\documentstyle[12pt]{article}

\newcommand{\be}{\begin{equation}}
\newcommand{\ee}{\end{equation}}
\newcommand{\bae}{\begin{eqnarray}}
\newcommand{\eae}{\end{eqnarray}}
\def\hR12{\hat{R}_{12}}
\def\inprod#1#2{\left\langle #1, #2\right\rangle}
\newcommand{\rtimes}{\mbox{$\times\!\rule{0.3pt}{1.1ex}\,$}}
\def\p{\partial}
\def\bb{\bibitem}
\def\q2{q^{2}}

\def\Hm{q^{-2H_{0}/n}}
\def\qi2{q^{-2}}
\def\l{\lambda}

\def\fe{& = &}

\def\ff{\nn \\}
\def\smash{{\A \rtimes \U}}
\def\px{\partial_{x}}
\def\py{\partial_{y}}
\def\lx{\L_{x}}
\def\ix{\inn_{x}}
\def\iy{\inn_{y}}
\def\GL{GL_{q}(n)}
\def\P{{\cal P}}
\def\A{{\cal A}}
\def\U{{\cal U}}
\def\B{{\cal B}}
\def\O{{\cal O}}
\def\R{\hat{R}}
\def\hRijkl{\R_{ij,kl}}
\def\inn{\mbox{\boldmath $i$}}

\def\delA{\Delta_{\A}}
\def\xj{x_{j}}

\def\xk{x_{k}}
\def\xl{x_{l}}

\def\iAij{(A^{-1})_{ij}}
\def\iAkl{(A^{-1})_{kl}}
\def\k2{x_{i}^{(2')}}
\def\Okrls{\O_{kr,ls}}
\def\Xkl{X_{kl}}
\def\Xsr{X_{sr}}
\def\iXkl{\inn_{\Xkl}}
\def\iXsr{\inn_{\Xsr}}
\def\men#1{~(\ref{#1})}
\def\pkl#1{\inprod{X_{kl}}{A^{-1}_{#1}}}
\def\ikl{\inn_{kl}}
\def\irs{\inn_{rs}}
\def\xbj{\bar{x}_{j}}
\def\hR12{\R_{12}}
\def\i1{\inn_{1}}
\def\i2{\inn_{2}}

\def\L{\mbox{\boldmath $\pounds$}}
\def\d{{\bf d}}
\def\dbf{{\bf d}}
\def\de{\mbox{\boldmath $d$}}
\def\nn{\nonumber}
\def\id{\mathop{\rm id}}
\def\uA{1_{\A}}
\def\uU{1_{\U}}

\def\uB{1_{\B}}
\def\tr{\triangleright}
\def\ra{\rightarrow}
\begin{document}
\begin{titlepage}
\begin{center}
\today     \hfill    LBL-35034 \\
          \hfill    UCB-PTH-94/01 \\

\vskip .3in

{\large \bf Induced Extended Calculus On The Quantum Plane}
\footnote{This work was supported in part by the Director, Office of
Energy Research, Office of High Energy and Nuclear Physics, Division of
High Energy Physics of the U.S. Department of Energy under Contract
DE-AC03-76SF00098 and in part by the National Science Foundation under
grant PHY90-21139.}

\vskip .3in

Chryssomalis Chryssomalakos\footnote[1]{e-mail address:
chryss@physics.berkeley.edu}, Peter Schupp and Bruno Zumino \\[.5in]
{\em  Department of Physics\\
      and\\
      Theoretical Physics Group\\
      Lawrence Berkeley Laboratory\\
      University of California at Berkeley\\
      Berkeley, California 94720}
\end{center}

\vskip .3in

\begin{abstract}
The non-commutative differential calculus on quantum groups can be
extended by introducing, in analogy with the classical case, inner
product operators and Lie derivatives. For the case of $\GL$ we
show how this extended calculus induces by coaction a similar
extended calculus, covariant under $\GL$, on the quantum plane. In
this way, inner product operators and Lie derivatives can be
introduced on the plane as well. The situation with other quantum
groups and quantum spaces is briefly discussed. Explicit formulas
are given for the two dimensional quantum plane.
\end{abstract}
\end{titlepage}
\renewcommand{\thepage}{\roman{page}}
\setcounter{page}{2}
\mbox{ }

\vskip 1in

\begin{center}
{\bf Disclaimer}
\end{center}

\vskip .2in

\begin{scriptsize}
\begin{quotation}
This document was prepared as an account of work sponsored by the United
States Government.  Neither the United States Government nor any agency
thereof, nor The Regents of the University of California, nor any of
their
employees, makes any warranty, express or implied, or assumes any legal
liability or responsibility for the accuracy, completeness, or
usefulness
of any information, apparatus, product, or process disclosed, or
represents
that its use would not infringe privately owned rights.  Reference
herein
to any specific commercial products process, or service by its trade
name,
trademark, manufacturer, or otherwise, does not necessarily
constitute or
imply its endorsement, recommendation, or favoring by the United States
Government or any agency thereof, or The Regents of the University of
California.  The views and opinions of authors expressed herein do not
necessarily state or reflect those of the United States
Government or any
agency thereof of The Regents of the University of California and shall
not be used for advertising or product endorsement purposes.
\end{quotation}
\end{scriptsize}

\vskip 2in

\begin{center}
\begin{small}
{\it Lawrence Berkeley Laboratory is an equal opportunity employer.}
\end{small}
\end{center}
\newpage
\renewcommand{\thepage}{\arabic{page}}
\setcounter{page}{1}
\section{Introduction}
The differential calculus on quantum groups~\cite{Woron1,Woron2}
involves functions on the group, differentials, differential forms
and derivatives. The basic differential operator $\d$ satisfies the
standard properties such as linearity, $(\d 1) =0, \, \d^{2}=0$ and
the undeformed Leibniz rule:
\be
(\d fg) = (\d f) g + (-1)^{k} f (\d g) \, , \label{unLei}
\ee
where $k$ is the degree of the form $f$. As shown explicitly for
$\GL$ in~\cite{SWZlinear}, this calculus can be extended by the
introduction of inner product operators $\inn_{X}$ in terms of
which Lie derivatives are given by the undeformed formula:
\be
\L_{X} = \d \inn_{X} + \inn_{X} \d  , \label{Ldefi}
\ee
valid for forms of any degree.

The covariant differential calculus on the quantum plane
similarly involves functions, differentials, differential forms and
derivatives. The differential operator $\de$ on the plane
satisfies again the standard undeformed properties such
as\men{unLei}. It is natural to ask whether the calculus on the
plane can also be extended by the introduction of inner product
operators and Lie derivatives related to each other by a formula
analogous to\men{Ldefi}. In this paper we show that this is indeed
possible and that the extended calculus on the plane can actually
be considered as being induced by the extended calculus on the
group $\GL$ under which the quantum plane is covariant.

For differentiable manifolds, the introduction of the inner product
operator and its relation\men{Ldefi} with the Lie derivative
provided a kind of algebraization of the differential calculus. We
find it pleasing that the same undeformed relations\men{unLei}
and\men{Ldefi} are valid in the more clearly algebraic context of
non-commutative calculus.

The extended calculus on the plane can also be established
independently, without reference to that on $\GL$. It represents an
appropriate combination of a bosonic and a fermionic quantum
calculus on the plane~\cite{PuszWor,Pusz,BZMPLA}, the
differentials being
fermionic and the inner product operators fermionic derivatives
with respect to the differentials. The formulas expressing
operations on the quantum group in terms of operations on the plane
provide then a realization of the former on the plane.

Most of this paper is concerned with the quantum group $\GL$ and
with the quantum plane covariant under its coaction. In this case,
the realization
 mentioned above is especially simple. For
instance, the basic vector fields of the quantum Lie algebra of
$\GL$ are realized by differential operators on the quantum
plane, linear in the coordinates and their corresponding
derivatives. For other quantum groups the realization is more
complicated and requires quantum pseudodifferential operators, {\em
i.e.} non-linear functions of the coordinates and the derivatives.
This is shown explicitly for $SL_{q}(n)$ in section 4. A general
procedure for obtaining such realizations, applicable to other
quantum groups, is also outlined.

In section 2 we collect some well known properties of Hopf
algebras, actions and coactions which we shall use later, mostly in
order to establish our notation. We also review the extended
differential calculus on quantum groups, with special emphasis on
$\GL$. In section 3 we use these results to work out the extended
calculus on the quantum plane. In section 4 we explain how, for
groups other than $\GL$, the vector fields are realized on the
corresponding quantum space by means of quantum pseudodifferential
operators. Finally, in the appendix, we give explicitly all
commutation relations of the extended calculus for the
two-dimensional quantum plane.
\section{Preliminaries}
We collect here, in the interest of self-containment, several
definitions and results that we will need later. Detailed
treatments of the topics touched upon here can be found in the
references.
\subsection{Hopf Algebras, Actions And Coactions}
We start with the definition of a
{\em Hopf algebra}~\cite{Abe,Majid1,Swee}. An associative
unital algebra $\A$ over a field $k$ is called a Hopf algebra if it
posesses a {\em coproduct} $\Delta : \A \ra \A \otimes \A$, a {\em
counit}
$\epsilon : \A \ra k$ and an {\em antipode} $S : \A \ra \A $ which
obey:
\bae
\Delta (ab)  & = & \Delta (a) \Delta (b)  \\
\epsilon (ab)  & = & \epsilon (a) \epsilon (b) \\
(\Delta \otimes \id ) \circ \Delta  & = & (\id \otimes \Delta ) \circ
\Delta \\
(\epsilon \otimes \id ) \circ \Delta  & = & \id \\
(\id \otimes \epsilon ) \circ \Delta  & = & \id \\
\cdot (S \otimes \id ) \circ \Delta (a)  & = & \epsilon (a) \uA \\
\cdot (\id \otimes S) \circ \Delta (a)  & = & \epsilon (a) \uA \\
\Delta (\uA )  & = & \uA \otimes \uA \\
\epsilon (\uA )  & = & 1.
\eae
We will often use the Sweedler notation for the coproduct:
\be
\Delta (a) \equiv \sum_{i} a_{(1)}^{i} \otimes a_{(2)}^{i} \equiv
a_{(1)} \otimes a_{(2)} \, .
\ee
Also:
\be
(\Delta \otimes \id ) \circ \Delta (a) \equiv a_{(1)} \otimes
a_{(2)} \otimes a_{(3)}
\ee
and so on for higher powers of the coproduct.

Two Hopf algebras $\U$ and $\A$ are said to be {\em dually paired} if
there exists a non-degenerate inner product $\inprod{\cdot}{\cdot} : \U
\otimes \A \ra k$ which relates the two Hopf structures as follows
($x,y \in \U, \, a,b \in \A$):
\bae
\inprod{x}{ab}  & = & \inprod{x_{(1)}}{a} \inprod{x_{(2)}}{b} \\
\inprod{xy}{a}  & = & \inprod{x}{a_{(1)}} \inprod{y}{a_{(2)}} \\
\inprod{\uU}{a}  & = & \epsilon (a) \\
\inprod{x}{\uA}  & = & \epsilon (x) \\
\inprod{S(x)}{a}  & = & \inprod{x}{S(a)} \, .
\eae
We will assume, as is the case in the applications we consider,
that $S^{-1}$ exists. An algebra that satisfies the Hopf algebra
conditions except those involving the antipode is called a
{\em bialgebra}. Given an algebra $\B$ and a bialgebra $\A$, we say
``$\A$ coacts from the right on $\B$'' if there exists a {\em
coaction} $\delA : \B \ra \B \otimes \A$ satisfying ($b,c \in \B$):
\bae
\delA (bc)  & = & \delA (b) \delA (c) \\
( \delA \otimes \id ) \circ \delA  & = & (\id \otimes \Delta ) \circ
\delA \\
(\id \otimes \epsilon ) \circ \delA (b)  & = & b \\
\delA (\uB )  & = & \uB \otimes \uA .
\eae
We will use a Sweedler-like notation for coactions:
\be
\delA (b) \equiv \sum_{i} b^{(1)}_{i} \otimes b^{(2')}_{i} \equiv
b^{(1)} \otimes b^{(2')}
\ee
where, we remind the reader, $b^{(1)} \otimes b^{(2')} \in \B
\otimes \A$.

An algebra $\A$ {\em acts from the left} on an algebra $\B$ if
there exists a map $\tr : \A \otimes \B \ra \B, \, a \otimes b
\mapsto a \tr b$, satisfying ($a,\, a' \in \A, \, b \in \B$):
\bae
(aa') \tr b  & = & a \tr ( a' \tr b ) \\
\uA \tr b  & = & b.
\eae
Right coactions of $\A$ on $\B$ give rise to actions from the left
of $\A^{\ast}$ (the dual of $\A$) on $\B$ according to ($a \in
\A^{\ast}, \, b \in \B$):
\be
a \tr b \equiv b^{(1)} \inprod{a}{b^{(2')}}.
\ee
Given two dually paired Hopf algebras $\U$ and $\A$, one can
construct a new algebra, their {\em semidirect product} $\smash$, in the
following way: as a vector space, $\smash$ is the tensor product of
$\A$ and $\U$. The product in $\smash$ is defined as ($a,b \in \A,
\, x,y \in \U$):
\bae
(a \otimes x) (b \otimes y)  & = & a (x_{(1)} \tr b) \otimes x_{(2)} y
\nn \\
 & \equiv & a b_{(1)} \inprod{x_{(1)}}{b_{(2)}} \otimes x_{(2)} y.
\label{semi1}
\eae
\subsection{Extended Calculus On Quantum Groups}
The concepts presented above are fully employed in the
description of Quantum
Groups~\cite{Dri,Woron1}. We examine here the
case of $\mbox{Fun}_{q}
(GL(n))$~\cite{RTF,SWZlinear}. Let
$\A$ be the algebra generated by the unit $\uA$ and the
elements $A_{ij}$ of an $n \times n$ matrix $A$ modulo the
relations:
\be
\hR12 A_{1} A_{2} = A_{1} A_{2} \hR12 \label{RAA}
\ee
The $n^{2} \times n^{2}$ matrix $\R$ is a solution of the quantum
Yang-Baxter equation:
\be
\hR12 \R_{23} \hR12 = \R_{23} \hR12 \R_{23} \label{QYBE}
\ee
and satisfies the characteristic equation:
\be
\R^{2} - \lambda \R - 1 = 0, \, \, \, \lambda \equiv q-q^{-1} \, .
\label{char1}
\ee
The explicit form of $\R$ is given by $\hRijkl = R_{ji,kl}$ where
$R$ is the $\GL$ $R$-matrix of~\cite{RTF}, given by:
\be
R= q \sum_{i} e_{ii} \otimes e_{ii} + \sum_{i \neq j} e_{ii}
\otimes e_{jj} + \lambda \sum_{i > j} e_{ij} \otimes e_{ji}
\ee
where $i,j = 1, \ldots ,n$ and $e_{ij}$ is the $n \times n$ matrix
with single nonzero element (equal to 1) at $(i,j)$. The above
algebra can be endowed with a Hopf structure via the definitions:
\bae
\Delta (A_{ij}) & = & A_{ik} \otimes A_{kj} \\
\epsilon (A_{ij}) & = & \delta_{ij} \\
S(A_{ij}) & = & (A^{-1})_{ij} \label{HopfA} \, .
\eae
We now construct $\U$ so that $\A$ and $\U$ are dually paired Hopf
algebras. $\U$ is generated by the unit $\uU$ and the elements
$X_{ij}$ of the matrix $X$ whose inner product with the generators
of $\A$ is given by:
\bae
\inprod{X_{1}}{A_{2}} & = & \lambda^{-1} (1- \R^{2})_{12} \label{pair}
\\
\inprod{X}{\uA} & = & 0
\eae
and satisfy:
\be
\R_{12} X_{2} \R_{12} X_{2} - X_{2} \R_{12} X_{2} \R_{12} = \l^{-1}
(\R_{12}^{2} X_{2} - X_{2} \R_{12}^{2}) \, . \label{XX}
\ee
Notice that for $\GL$, the right hand side of\men{pair} simplifies
to $-\R_{12}$ due to the characteristic equation\men{char1} - we
leave it as is though since this form is valid for other quantum groups
as well.

The coproduct of $X_{ij}$ is of the form:
\be
\Delta (X_{ij}) = X_{ij} \otimes \uU + \O_{ij,kl} \otimes X_{kl}
\label{copX}
\ee
(with $\O_{ij,kl} \in \U$ typically non-linear in the $X_{ij}$'s) and
 gives rise to the following $X - A$ commutation relations in
$\smash$:
\be
X_{1} A_{2} = A_{2} \R_{12} X_{2} \R_{12} + \lambda^{-1} A_{2}
(1-\R^{2})_{12} \, . \label{XA}
\ee
One can now introduce an (undeformed) exterior derivative $\d$
which maps $k$-forms to $k+1$-forms (functions being 0-forms) and
satisfies:
\bae
\d^{2} & = & 0 \\
(\d fg) & = & (\d f) g +(-1)^{k} f (\d g) \nn \\
 & \equiv & df \; g + (-1)^{k} f \; dg \label{extder}
\eae
where $g$ is any form and $f$ is a $k$-form \footnote{$(\d f)$ is
the action of $\d$ on $f$ (the differential of $f$), denoted by
$df$ (to be read as a single symbol)}. The matrix $\Omega$ of
Cartan-Maurer forms is given by~\cite{SWZlinear,BZLeipzig}:
\be
\Omega \equiv A^{-1} dA = -dA^{-1} \; A \label{omegadef}
\ee
and satisfies:
\bae
\Omega_{1} A_{2} \fe A_{2} \R_{12}^{-1} \Omega_{2} \R_{12}^{-1}
\label{omegaA} \\
\Omega_{1} dA_{2} \fe -dA_{2} \R^{-1}_{12} \Omega_{2} \R_{12} \, .
\label{omegadA}
\eae
Inner derivations $\inn_{X}$ and Lie derivatives ${\bf \pounds}_{X}$
complete the construction. They obey, among other, the relations:
\bae
\inn_{\uU} & = & 0 \\
\inn_{X_{1}} A_{2} & = & ((X_{1})_{(1)} \tr A_{2} ) \inn_{(X_{1})_{(2)}}
\nn \\
 & = & A_{2} \R_{12} \inn_{X_{2}} \R_{12} \\ \label{iA}
\inn_{X_{1}} dA_{2} & = & X_{1} \tr A_{2} - d((X_{1})_{(1)} \tr A_{2}
) \inn_{(X_{1})_{(2)}} \label{idA} \\
{\bf \pounds}_{X} & \equiv & \inn_{X} \d + \d \inn_{X} \label{Ldef1} \\
\L_{X} \d \fe \d \L_{X} \, .
\eae
Using $\Omega$ instead of $dA$ and substituting explicit
expressions for the inner products that occur in\men{idA} we get:
\be
\R_{12} \inn_{X_{2}} \R_{12} \Omega_{2} = -\Omega_{2} \R_{12}
\inn_{X_{2}} \R_{12} + \l^{-1}(1-\R^{2})_{12} \, . \label{iomega}
\ee
We undertake now the task of letting the structure described above
induce a corresponding one on the quantum plane - this is the
subject of the next section.

\section{Extended Calculus On The Quantum Plane}
We work in the following with the (non-commutative) algebra $\P$ of
functions on the quantum plane, generated by the unit $1_{\P}$ and
the coordinates $\xj ,\, j=1,\ldots ,n$ which satisfy:
\be
x_{2} x_{1} = q^{-1} \hR12 x_{2} x_{1} \label{xx}
\ee
(notice that the above can be obtained from the ``standard'' quantum
plane of~\cite{PuszWor,WZ} by letting $q \rightarrow q^{-1}$ and
remembering that
$\R_{12}(q^{-1})=\R_{21}^{-1}(q)$). The algebra $\P$ admits the
right coaction $\delA :\P \rightarrow \P \otimes \A$ (with $\A
\equiv \mbox{Fun}_{q} (GL(n))$ defined on the generators $\xk$ by:
\be
\delA (\xk)=\xl \otimes \iAkl \label{rcoact1}
\ee
($A$ is a $\GL$ matrix) and extended to the whole of $\P$
multiplicatively - we take of course $\delA (1_{\P})=1_{\P} \otimes
1_{\A}$. Our next step will be to give commutation relations
between elements of $\A \rtimes \U$ and $\P$. To accomplish this,
notice that~(\ref{rcoact1}) allows us to embed $\P$ in $\P \otimes
\smash$ (since $\smash$ contains $\A$ as a subalgebra) via $x_{i}
\mapsto x_{i}^{(1)} \otimes \k2$. Then, the trivial embeding $\smash
\rightarrow 1_{\P} \otimes \smash$ allows us to compute, for
example, commutation relations between elements of $\U$ and $\P$
($\chi \in \U$ ):
\bae
\chi x_{i} & \mapsto & (1 \otimes \chi) \delA (x_{i}) \nn \\
 & = & x_{i}^{(1)} \otimes (\k2 )_{(1)}
\inprod{\chi_{(1)}}{(\k2 )_{(2)}}
\chi_{(2)} \nn \\
 & = & (x_{i}^{(1)})^{(1)} \otimes (x_{i}^{(1)})^{(2')}
\inprod{\chi_{(1)}}{\k2} \chi_{(2)} \nn \\
 & = & \delA(x_{i}^{(1)}) \inprod{\chi_{(1)}}{\k2} (1 \otimes \chi_{(2)}
) \nn \\
\Rightarrow \chi x_{i} & = & x_{i}^{(1)} \inprod{\chi_{(1)}}{\k2}
\chi_{(2)}.
\eae
We will now specify $\chi$ to be one of the bicovariant generators
$\Xkl$ introduced in the previous section.
 Using\men{copX} for their coproduct and remembering that
$\inn_{1_{\U}}=0$, we find:
\be
\iXkl x_{i} = x_{i}^{(1)} \inprod{\Okrls \; }{\k2}
\iXsr \, . \label{ix1}
\ee
To facilitate upcoming computations, we introduce now a simplified
notation as follows. We write for the coaction of $x_{i}$:
\be
x_{i} = \iAij \xbj
\ee
where $\xbj$ are some ``fixed'' coordinates on the plane and the
elements of $A$ are taken to commute with them. To compute
explicitly the relations implied by\men{ix1} we start from the $\inn
 -A$ commutation relations of\men{iA} to get:
\bae
\inn_{1} A_{2} & = & A_{2} \hR12 \inn_{2} \hR12 \nonumber \\
\Rightarrow A^{-1}_{2} \inn_{1} & = & \hR12 \inn_{2} \hR12 A^{-1}_{2}
\, .
\eae
Multiplying now from the right by $\bar{x}_{2}$ (which commutes with
$\inn_{1}$ as well) we find:
\be
x_{2} \inn_{1} = \hR12 \i2 \hR12 x_{2} \label{ix2} \, .
\ee
In the same spirit, we introduce the differentials $dx_{i}$ via:
\be
dx_{i}=d(\iAij ) \xbj = -\Omega_{ij} x_{j} \, .  \label{diffdef}
\ee
With the help of\men{omegaA}, this implies:
\be
(dx_{2} ) x_{1} = q \hR12 x_{2} (dx_{1} ) \label{xdx}
\ee
while\men{omegadA} similarly induces:
\be
dx_{2} dx_{1} = -q \R_{12} dx_{2} dx_{1} \, . \label{dxdx}
\ee
On the other hand, using the general relation\men{idA} we get
($\ikl \equiv \iXkl$ ):
\be
\ikl dx_{i} = \xj \pkl{ij} - dx_{j} \inprod{\Okrls}{\iAij} \irs
 \, .\label{idx}
\ee
Considering the first term in\men{idx} we would like now to realize
the inner derivations $\inn_{kl}$ of the quantum group in terms of
inner derivations $\inn_{k}$ on the plane via:
\bae
\ikl & \sim & x_{m} \pkl{nm} \inn_{n} \nonumber \\
 & \equiv &  Q_{kn,ml} \; x_{m} \; \inn_{n}  \, .\label{irep}
\eae
Substitution in\men{idx} gives:
\bae
Q_{kr,sl} \; x_{s} \; \inn_{r} \; dx_{i} & = & Q_{ki,jl} x_{j}
- Q_{rm,ns}
\inprod{\O_{kl,rs}}{\iAij} dx_{j} \;  x_{n} \; \inn_{m}
 \, .\label{ugly1}
\eae
We now attempt to extract $\inn_{k} - dx_{l}$ commutation relations
from this equation. To succeed in this, it will be necessary to
resort to explicit numerical computation that will parallel the
above. We start from\men{iomega} and use\men{diffdef} to get:
\be
\i2 \hR12 (dx_{2} ) = x_{2} - \hR12^{-1} (dx_{2} ) \inn_{1} \label{idx1}
\ee
(this is the explicit form of\men{idx}). It is time now for some
$R$-matrix trickery. Written out explicitly in terms of
indices,\men{idx1} gives:
\be
\inn_{kk'} R_{k'i,jk''} \; dx_{k''} + \hat{R}^{-1}_{ik,i'k'} \; dx_{k'}
\; \inn_{i'j} = x_{k} \; \delta_{ij}  \, . \label{idx2}
\ee
However, the $R$-matrix satisfies:
\bae
R_{12}^{t_{1}} & = & D_{1} \tilde{R}_{12} D_{1}^{-1} \; \;
\mbox{or:} \nonumber \\
  R_{k'i,jk''} & = & D_{j} D_{k'}^{-1}
\tilde{R}_{ji,k'k''}
\eae
where:
\be
D_{ij} \equiv D_{i} \delta_{ij} = \mbox{diag} (q^{-2n+1},
q^{-2n+3}, \ldots ,q^{-1})
\ee
and $\tilde{R} \equiv ((R^{-1})^{t_{1}})^{-1}$ (with $t_{1}$
denoting transposition in the first space). Using this in\men{idx2}
we get:
\bae
D_{j} D_{k'}^{-1} \tilde{R}_{ji,k'k''} \; \inn_{kk'} \; dx_{k''} & = &
x_{k} \; \delta_{ij} - \R^{-1}_{ik,i'k'} \; dx_{k'} \; \inn_{i'j}
\Rightarrow \nonumber \\
\Rightarrow D_{p}^{-1} \; \inn_{kp} \; dx_{q} & = & -D_{j}^{-1}
(R^{-1})^{t_{1}}_{pq,ji} (R^{-1})^{t_{1}}_{ik,i'k'} \; dx_{k'} \;
\inn_{i'j} \nonumber \\
 & & +D_{i}^{-1} (R^{-1})^{t_{1}}_{pq,ii} \; x_{k} \, . \label{idx3}
\eae
A second property of $R$ comes now to our help:
\be
D_{i}^{-1} R^{-1}_{iq,pi} = \delta_{qp}
\ee
so that\men{idx3} becomes:
\bae
D_{p}^{-1} \inn_{kp} dx_{q} & = & -D_{j}^{-1}
(R^{-1})^{t_{1}}_{pq,ji} (\R^{-1})_{ik,i'k'} \; dx_{k'} \; \inn_{i'j}
 +\delta_{qp} \; x_{k} \, . \label{idx4}
\eae
Before realizing $\ikl$ according to\men{irep} we extract the
inner product that appears in\men{irep} from the $X - A$ commutation
relations\men{XA}. Using the characteristic equation for $\R$, it
becomes:
\be
\pkl{nm} = \delta_{km} D_{ln} , \label{pair2}
\ee
so that the realization of $\ikl$ on the plane is given by:
\be
\ikl \sim D_{ls} x_{k} \inn_{s} \, . \label{irep2}
\ee
Substituting this in\men{idx4} we find:
\bae
x_{k} \; \inn_{p} \; dx_{q} & = & -D_{j}^{-1} (R^{-1})^{t_{1}}_{pq,ji}
(\R^{-1})_{ik,i'k'} \; dx_{k'} \; x_{i'} \; D_{j} \; \inn_{j}
\nonumber \\
 & & +x_{k} \; \delta_{qp}
\eae
or, with the help of\men{xdx}:
\be
x_{k}\; (\inn_{p} \; dx_{q} + q \R^{-1}_{jq,ip} \; dx_{i} \; \inn_{j} -
\delta_{qp}) = 0.
\ee
The factor in perentheses supplies us with $\inn_{p}- dx_{q}$
commutation relations:
\be
\inn_{p} \; dx_{q} = \delta_{pq} - q\R^{-1}_{jq,ip} \; dx_{i} \;
\inn_{j} .
\label{ipdx}
\ee
A similar computation, starting from\men{iA}, gives:
\be
\inn_{p} \; x_{q} = q^{-1} \R^{-1}_{jq,ip} \; x_{i} \; \inn_{j} \,
. \label{ipx}
\ee
We wish now to include partial
derivatives $\partial_{r}$ with respect to the coordinates $x_{r}$
as well as Lie derivatives $\pounds_{s}$. The former we
introduce by realizing the vector fields $X_{kl}$ via:
\bae
X_{kl} & \sim & \pkl{rs} x_{s} \partial_{r} \nn \\
 & = & D_{lr} \; x_{k} \partial_{r} \, . \label{Xrep}
\eae
Starting from:
\be
D_{p}^{-1} X_{kp} \; x_{q} = x_{k} \; \delta_{qp} + D_{j}^{-1}
(R^{-1})_{jq,pi} (\R^{-1})_{ik,i'k'} \; x_{k'} \; X_{i'j}
\ee
(easily derivable from the $X-A$ commutation relations), we obtain:
\be
\partial_{p} x_{q} = \delta_{qp} + q^{-1} \R^{-1}_{jq,ip} \; x_{i}
\partial_{j} \label{derx}
\ee
and similarly, making use of the characteristic equation for $\R$:
\be
\partial_{p} \; dx_{q} = q \R_{jq,ip} \; dx_{i} \; \partial_{j}
\, .
\label{derdx}
\ee
On the other hand,\men{derx} and the $X-X$ commutation relations
give:
\be
\partial_{k} \partial_{l} = q^{-1} \R_{pq,kl} \partial_{p}
\partial_{q} \, . \label{derder}
\ee
An exterior derivative $\de$ on the plane is given by:
\be
\de \equiv dx_{i} \partial_{i} \, ; \label{delta}
\ee
it realizes the action of $\d$ on the plane implied
by\men{diffdef} so that it satisfies relations analogous
to\men{extder}:
\bae
\de^{2} \fe 0  \\
\de \alpha \fe d\alpha +(-1)^{k} \alpha \de \label{extderp}
\eae
where $\alpha$ is a $k$-form on the plane - in particular: $\de
x_{i}= dx_{i} + x_{i} \de$.

We may now complete the extended calculus structure on the quantum
plane by introducing the Lie derivatives $\pounds_{s}$, in the
canonical way, through:
\be
\pounds_{s} \equiv \inn_{s} \de + \de \, \inn_{s} \, . \label{Lpdef}
\ee
These enter naturally in the realization of $\pounds_{kl}$
($\equiv \L_{X_{kl}}$):
\bae
\L_{kl} & = & \ikl \dbf +\dbf \; \ikl \nonumber \\
 & \sim & x_{k} \; D_{ls} \; \inn_{s}\; \de + \de x_{k} \; D_{ls} \;
\inn_{s} \nonumber
\\
 & = & x_{k} \; D_{ls} \; \inn_{s} \; \de + dx_{k} \;  D_{ls} \;
\inn_{s} + x_{k} \;
D_{ls}\; \de \; \inn_{s} \nonumber \\
\Rightarrow \L_{kl} & \sim & x_{k} \; D_{ls} \; \L_{s} + dx_{k} \;
D_{ls} \; \inn_{s} . \label{Lrep}
\eae
Their commutation relations with the coordinates
$x_{i}$ are now easily computed:
\bae
\L_{p} \; x_{q} & = & (\de \; \inn_{p} + \inn_{p} \; \de) x_{q} \nn \\
 & = & \de q^{-1} (\R^{-1})_{jq,ip} \; x_{i} \; \inn_{j} + \inn_{p}
dx_{q} + \inn_{p} \;
x_{q} \de
\nn \\
 & = & q^{-1} (\R^{-1})_{jq,ip} \; (dx_{i}+x_{i} \de )\;  \inn_{j} +
\delta_{pq} - q(\R^{-1})_{jq,ip} \; dx_{i} \; \inn_{j} \nn \\
 & & + q^{-1} (\R^{-1})_{jq,ip} \; x_{i} \; \inn_{j} \; \de
\Rightarrow \nn
\\
\Rightarrow \L_{p} \; x_{q} & = & \delta_{pq} + q^{-1}
(\R^{-1})_{jq,ip} \; x_{i} \; \L_{j} - \lambda (\R^{-1})_{jq,ip} \;
dx_{i} \;
\inn_{j} . \label{Lpx}
\eae
Notice the appearance of a term, absent in the classical
case, which is bilinear in the differentials and the inner derivations.
Proceeding similarly, we find:
\be
\L_{p} \; dx_{q} = q (\R^{-1})_{jq,ip} \; dx_{i} \; \L_{j} \, .
\label{Lpdx}
\ee
The following, easily verifiable, identities are often useful in
computations:
\bae
\de \L_{p} \fe \L_{p} \de \\
\de \partial_{p} \fe q^{-2} \partial_{p} \de \\
\partial_{p} \fe q^{2} \de \inn_{p} + \inn_{p} \de \, .
\eae
To complete the description of the above scheme, we give below the
remaining commutation relations among the generators of the
extended calculus:
\bae
\L_{p} \p_{q} \fe q^{-1} \R_{ji,pq} \p_{j} \L_{i} \\
\inn_{p} \p_{q} \fe q \R_{ji,pq} \p_{j} \inn_{i} \\
\L_{p} \L_{q} \fe q^{-1} \R_{ji,pq} \L_{j} \L_{i} \\
\L_{p} \inn_{q} \fe q^{-1} \R_{ji,pq} \inn_{j} \L_{i} \\
\inn_{p} \inn_{q} \fe -q \R_{ji,pq} \inn_{j} \inn_{i} \, .
\eae
The explicit form, for $n=2$, of this extended calculus is given in
the appendix.
\section{Realizations Via Pseudodifferential Operators}
We wish now to remark briefly on the case of other quantum groups.
We start with $SL_{q}(n)$ and follow the approach and conventions
of~\cite{SWZlinear}. It will be convenient in the following to
introduce the matrix $Y$ of vector fields~\cite{RSTS,SWZalg} on $\GL$
through:
\be
Y=1-\l X \, .  \label{Y}
\ee
One can define a determinant-like quantity $DetY \equiv
q^{2H_{0}}$, the precise definition of which can be found
in~\cite{SWZalg,BZSala} , which is homogeneous of degree $n$ in the
$Y_{ij}$'s and commutes with them. Vector fields $V$ on the group
manifold of $SL_{q}(n)$ can now be defined via:
\be
1-\l V \equiv Z \equiv \Hm Y = \Hm (1-\l X)  \, . \label{VZYX}
\ee
The determinant of $Z$, using the same definition as for that of
$Y$, is found to be equal to 1 and this restricts the number of
generators to $n^{2} -1$. Non-homogeneous constraint equations like
this, do not allow realizations of the action of the $V_{ij}$'s
on the plane bilinear in the coordinates and the derivatives.
However,\men{VZYX} offers a realization in terms of
pseudodifferential operators. We will need the relation:
\be
q^{2H_{0}} A = \q2 A q^{2H_{0}} \label{DetYA}
\ee to make this precise. Indeed, the above equation implies:
\be
q^{2H_{0}} x_{i} = \qi2 x_{i} q^{2H_{0}}  \, . \label{DetYx}
\ee
On the other hand, one easily verifies that:
\be
(1 - q^{-1} \l x \cdot \partial ) x_{i} = \qi2 x_{i} (1- q^{-1} \l
x \cdot \partial ) \label{rescale}
\ee
which permits the realization:
\bae
V_{ij} \fe \l^{-1} (1- \Hm ) \delta_{ij} + \Hm X_{ij} \ff
 & \sim & \l^{-1} (1-(1-q^{-1} \l x \cdot \partial)^{-\frac{1}{n}})
\delta_{ij} \nn \\
 & & + (1-q^{-1} \l x \cdot \partial )^{-\frac{1}{n}} x_{i}
D_{jr} \partial_{r}   . \label{repV}
\eae
The above approach clearly relies on the fact that $V_{ij}$ can be
obtained in terms of the unconstrained $X_{ij}$ which admit simple
bilinear realizations. Such a construction is not known for
other quantum groups like, for example, $SO_{q}(n)$. Nevertheless,
it is not hard to outline a general procedure for obtaining
realizations, valid for those groups as well. We will use
$SL_{q}(2)$ as an illustrative example. In this case, starting
from\men{XA}, one can compute the action of the $V$'s on monomials
in the coordinates, ordered according to some standard ordering. We
use the notation:
\be
V = \left( \begin{array}{cc}
v_{1} & v_{+} \\
v_{-} & v_{2}
\end{array} \right)  ,
\; \; \; x_{1} \equiv x, \;  x_{2} \equiv y \label{notation}
\ee
and find, for example:
\be
v_{-} \tr x^{k} y^{l} = q^{-2-l} [k]_{q} x^{k-1} y^{l+1}
\label{viminus}
\ee
where $[k]_{q} \equiv (1-q^{2k})/(1-q^{2})$. Regard, for the moment,
the above equation as referring to classical variables. Then it is
easy to realize $v_{-}$ in terms of classical coordinates and
derivatives:
\be
v_{-} \sim y' \partial_{x}' q^{-2-L} \frac{[K]_{q}}{K}
\label{vmrepc}
\ee
where $K \equiv x' \partial_{x}'$, $L \equiv y' \partial_{y}'$
and the primes are there to remind us that we are dealing with
classical quantities. It is well known that quantum
derivatives (acting on quantum coordinates) can be expressed in
terms of classical pseudodifferential operators (acting on the
corresponding classical coordinates) via invertible maps~\cite{Ogie}.
For example, in the case we are considering:
\be
\partial_{x} \sim \partial_{x}' \frac{[K]_{q^{-1}}}{K} q^{-2L}, \;
\; \;
  \partial_{y} \sim \partial_{y}' \frac{[L]_{q^{-1}}}{L} q^{-K}
\, .
\label{maps}
\ee
Inverting these and substituting in\men{vmrepc} we find, in
agreement with\men{repV}:
\be
v_{-} \sim q^{-3} (1-q^{-1} \l (x \partial_{x} +y \py ))^{-1/2} y
\px \, .\label{vmrepq}
\ee
Maps like the above are known for $SL_{q}(n)$ and
$SO_{q}(n)$~\cite{Ogie}. Clearly then, a computation along the same
lines can provide realizations for the Lie algebra
generators of them as well.
\section*{Aknowledgements}
We would like to thank Paul Watts for many helpful discussions.

This work was supported in part by the Director, Office of
Energy Research, Office of High Energy and Nuclear Physics, Division of
High Energy Physics of the U.S. Department of Energy under Contract
DE-AC03-76SF00098 and in part by the National Science Foundation under
grant PHY90-21139.

It is a pleasure to dedicate this paper to Ludwig D. Faddeev
\pagebreak
\appendix
\section{Extended Calculus On The 2-D Quantum Plane}
The $\R$-matrix for $\GL$ is:
\be
\R = \left( \begin{array}{cccc}
q & 0 & 0 & 0 \\
0 & \l & 1 & 0 \\
0 & 1 & 0 & 0 \\
0 & 0 & 0 & q
\end{array} \right), \; \; \;  \l \equiv q-q^{-1} . \nn
\ee
Using this, and the definitions:
\be
\xi \equiv \de x -x \de , \; \; \; \eta \equiv \de y -y \de
\ee
(with $\de \xi + \xi \de = \de \eta + \eta \de = 0$),
 the general formulas of section 3 give (we remind the
reader that the calculus presented here differs from that
of~\cite{WZ} by the substitution $q \mapsto q^{-1}$):
\bae
xy \fe q^{-1}yx \ff
& & \ff
x\xi \fe \qi2 \xi x \ff
x \eta \fe q^{-1}\eta x - q^{-1}\l \xi y \ff
y \xi \fe q^{-1}\xi y \ff
y \eta \fe \qi2 \eta y \ff
& & \ff
\px x \fe 1 + \qi2 x \px - q^{-1}\l y \py \ff
\px y \fe q^{-1}y \px \ff
\py x \fe q^{-1}x \py \ff
\py y \fe 1 + \qi2 y \py \ff
 & & \ff
 & & \ff
\pagebreak
\lx x \fe 1 + \qi2 x \lx - q^{-1}\l y \L_{y} - q^{-1}\l \xi \ix + \l^{2}
\eta \iy \ff
\lx y \fe q^{-1}y \lx - \l \eta \ix \ff
\L_{y} x \fe q^{-1}x \L_{y} - \l \xi \iy \ff
\L_{y} y \fe 1 + \qi2 y \L_{y} - q^{-1}\l \eta \iy \ff
 & & \ff
\ix x \fe \qi2 x \ix - q^{-1}\l y \iy \ff
\ix y \fe q^{-1} y \ix \ff
\iy x \fe q^{-1}x \iy \ff
\iy y \fe \qi2 y \iy \ff
 & & \ff
\xi \xi \fe 0 \ff
\eta \eta \fe 0 \ff
\xi \eta \fe -q \eta \xi \ff
 & & \ff
\px \xi \fe \q2 \xi \px \ff
\px \eta \fe q \eta \px \ff
\py \xi \fe q \xi \py \ff
\py \eta \fe \q2 \eta \py + q \l \xi \px \ff
 & & \ff
\lx \xi \fe \xi \lx - q \l \eta \L_{y} \ff
\lx \eta \fe q \eta \lx \ff
\L_{y} \xi \fe q \xi \L_{y} \ff
\L_{y} \eta \fe \eta \L_{y} \ff
 & & \ff
\ix \xi \fe 1 - \xi \ix + q \l \eta \iy \ff
\ix \eta \fe - q \eta \ix \ff
\iy \xi \fe -q \xi \iy \ff
\iy \eta \fe 1 - \eta \iy \ff
 & & \ff
\pagebreak
\px \py \fe q \py \px \ff
 & & \ff
\lx \px \fe \px \lx \ff
\lx \py \fe q^{-1}\py \lx + q^{-1}\l \px \L_{y} \ff
\L_{y} \px \fe q^{-1}\px \L_{y} \ff
\L_{y} \py \fe \py \L_{y} \ff
 & & \ff
\ix \px \fe \q2 \px \ix \ff
\ix \py \fe q \py \ix + q \l \px \iy \ff
\iy \px \fe q \px \iy \ff
\iy \py \fe \q2 \py \iy \ff
 & & \ff
\lx \L_{y} \fe q \L_{y} \lx \ff
 & & \ff
\lx \ix \fe \ix \lx \ff
\lx \iy \fe q^{-1}\iy \lx + q^{-1}\l \ix \L_{y} \ff
\L_{y} \ix \fe q^{-1}\ix \L_{y} \ff
\L_{y} \iy \fe \iy \L_{y} \ff
 & & \ff
\ix \ix \fe 0 \ff
\iy \iy \fe 0 \ff
\ix \iy \fe - q^{-1}\iy \ix \nn
\eae
\pagebreak


\begin{thebibliography}{abc}
\bb{Abe} E. Abe, \underline{``Hopf Algebras''}, Cambridge Univ.
Press, 1980
\bb{Dri} V. G. Drinfel'd, ``Quantum Groups'', in A. Gleason (ed.),
Proceedings of the ICM, Rhode Island (1987), AMS
\bb{Majid1} S. Majid,  ``Quasitriangular Hopf Algebras And
Yang-Baxter Equations'', {\em Int. J. Mod. Phys.} {\bf A5} 1-91 (1990)
\bb{Ogie} O. Ogievetsky, ``Differential Operators On Quantum Spaces
For $\GL$ And $SO_{q}(n)$'', {\em Lett. Math. Phys.} {\bf 24}
245-255 (1992)
\bb{PuszWor} W. Pusz and S. L. Woronowicz, ``Twisted Second
Quantization'', {\em Rep. Math.
Phys.} {\bf 27} 231-253 (1989)
\bb{Pusz} W. Pusz, ``Twisted Canonical
Anticommutation Relations'', {\em Rep. Math. Phys.} {\bf 27}
349-360 (1989)
\bb{RTF} N. Yu. Reshetikhin, L. A. Takhtadzhyan and L. D. Faddeev,
``Quantization Of Lie Groups And Lie Algebras'', {\em Leningrad
Math. J.} {\bf 1} 193-225 (1990)
\bibitem{RSTS} N. Yu. Reshetikhin and M. A. Semenov-Tian-Shansky,
``Quantum $R$ Matrices And Factorization Problems'', {\em J. Geom.
Phys.} {\bf 5} 533-550 (1988)
\bb{SWZlinear} P. Schupp, P. Watts and B. Zumino, ``Differential
Geometry On Linear Quantum Groups'', {\em Lett. Math. Phys.} {\bf
25} 139-147 (1992)
\bb{SWZalg} P. Schupp, P. Watts and B. Zumino, ``Bicovariant
Quantum Algebras And Quantum Lie Algebras'', {\em Commun. Math.
Phys.} {\bf 157}, 305-329 (1993)
\bb{Swee} M. E. Sweedler, \underline{``Hopf Algebras''}, Benjamin
(1969)
\bb{WZ} J. Wess and B. Zumino, ``Covariant Differential Calculus On
The Quantum Hyperplane'', {\em Nucl. Phys. B} (Proc. Suppl.) {\bf
18B} 302-312 (1990)
\bb{Woron1} S. L. Woronowicz, ``Compact Matrix Pseudogroups'', {\em
Commun. Math. Phys.} {\bf 111} 613-665 (1987)
\bb{Woron2} S. L. Woronowicz, ``Differential Calculus On Compact
Matrix Pseudogroups (Quantum Groups)'', {\em Commun. Math. Phys.}
{\bf 122} 125-170 (1989)
\bb{BZMPLA} B. Zumino, ``Deformation Of The Quantum Mechanical
Phase Space With Bosonic Or Fermionic Coordinates'', {\em Mod.
Phys. Lett.} {\bf A6}, 1225-1235 (1991)
\bb{BZSala} B. Zumino, ``Differential Calculus On Quantum Spaces
And Quantum Groups'', in M.O., M.S., J.M.G. (eds.), Proc.
XIX-th ICGTMP Salamanca (1992), CIEMAT/RSEF Madrid (1993)
\bb{BZLeipzig} B. Zumino, ``Introduction To The Differential
Geometry Of Quantum Groups'', in K. Schm\"{u}dgen (ed.), Math.
Phys. X, Proc. X-th IAMP Conf. Leipzig 1991, Heidelberg,
 Springer (1992)
\end{thebibliography}
\end{document}